# A COLLISION OF SUBCLUSTERS IN ABELL 754

ANN I. ZABLUDOFF

Observatories of the Carnegie Institution of Washington
813 Santa Barbara Street, Pasadena, CA, 91101, E-mail: aiz@libra.ociw.edu

and

DENNIS ZARITSKY

UCO/Lick Observatory and Board of Astronomy and Astrophysics,
University of California, Santa Cruz, CA, 95064, E-mail: dennis@lick.ucsc.edu

## ABSTRACT

We present direct evidence of a collision of subclusters in the galaxy cluster Abell 754. Our comparison of new optical data and archival ROSAT PSPC X-ray data reveal three collision signatures predicted by n-body/hydrodynamical simulations of hierarchical cluster evolution. First, there is strong evidence of a non-hydrostatic process; neither of the two major clumps in the galaxy distribution lies on the off-center peak of the X-ray emission from the intracluster gas. Second, the peak of the X-ray emission is elongated perpendicular to the collision axis defined by the centroids of the two galaxy clumps. Third, there is evidence of compression-heated gas; one of A754's two X-ray temperature components (Henry & Briel 1995) is among the hottest observed in any cluster and hotter than that inferred from the velocity dispersion of the associated galaxy clump. These signatures are consistent with the qualitative features of simulations (Evrard 1990a,b) in which two subclusters have collided in the plane of the sky during the last $\sim$Gyr. The detection of such collisions is crucial for understanding both the dynamics of individual clusters and the underlying cosmology. First, for systems like A754, estimating the cluster X-ray mass from assumptions of hydrostatic equilibrium and isothermality is incorrect and may produce the discrepancies sometimes found between X-ray masses and those derived from gravitational lens models (Babul & Miralda-Escudé 1994). Second, the fraction of nearby clusters in which subclusters have collided in the last $\sim$Gyr is especially sensitive to the mean mass density parameter $\Omega_0$ (cf. Richstone *et al.* 1992; Evrard *et al.* 1993; Lacey & Cole 1993). As we show for A754, we now have the means to identify recent collisions. With a large, well-defined cluster sample, it will be possible to place a new and powerful constraint on cosmological models.

## 1. INTRODUCTION

The complex dynamics of the gas, galaxies, and dark matter in clusters of galaxies provides clues to cluster evolution and to the underlying cosmology of the universe. In hierarchical models of the formation of large-scale structure, clusters evolve by accreting other clusters or smaller groups of galaxies from the field. Simulations show that when these systems collide, the effects can be observed for $\sim 1h^{-1}$ Gyr  $h$ is the Hubble constant in units of 100 km s$^{-1}$ Mpc$^{-1}$Evrard 1990b).

One key collision signature is the existence of hot, X-ray emitting, intracluster gas that is not in hydrostatic equilibrium with the cluster potential. During a collision of two subclusters, the peaks of the X-ray emission and the galaxy surface density will not coincide because the gas, unlike the galaxies and dark matter, is collisional. Although the gross features of the optical and X-ray data agree in a few well-studied clusters (Fabricant *et al.* 1986, 1993), these samples lack the spatial resolution to test the critical scales $< 0.5h^{-1}$ Mpc where simulations predict differences. Today, multi-object spectroscopy and a new generation of X-ray telescopes enable us to compare the spatial distribution of hundreds of cluster galaxies with high-resolution X-ray surface brightness maps to test the assumption of hydrostatic equilibrium.

Simulations show that the collision of the intracluster media of the subclusters produces two additional signatures (Evrard 1990b, Roettiger *et al.* 1993). First, the gas associated with a subcluster of galaxies is compressed in the direction of the collision, and if dense enough to be observed, appears as an elongated peak in the X-ray surface brightness distribution. Second, the collision significantly heats and may even shock some of the gas, creating an anomalously hot component in the spatially-resolved X-ray temperature map. The combination of high-sensitivity ROSAT PSPC data and multi-object spectroscopy allows us to test these predictions and to understand the history of recent collisions in clusters for the first time.

We present the results from the first cluster we have examined, Abell 754. There is evidence of a non-hydrostatic process; neither of the two major clumps in the galaxy distribution coincides with the off-center peak of the X-ray emission. We also observe the other collision signatures. The X-ray peak is distorted in a direction perpendicular to the axis on which the galaxy clumps lie, and the temperature of the X-ray gas surrounding one clump is a factor of two higher than that inferred from the clump's velocity dispersion.

We discuss the data briefly in §2 and the observed collision signatures in §3. We stress that deriving an X-ray cluster mass for a system that is neither isothermal nor in hydrostatic equilibrium is misleading. Most importantly, the detection of these signatures indicates that a collision has occurred in A754 within the last $\sim 1h^{-1}$ Gyr. A determination of the fraction of clusters with recent collisions, a more direct constraint of cosmological models than the fraction of clusters with substructure, is now within our grasp.

## 2. THE DATA

As part of a larger study of the internal dynamics of nearby clusters (Zabludoff 1995), we have obtained galaxy spectra and X-ray images for A754, a rich cluster at a mean redshift of 0.055 with an X-ray luminosity of $L_X = 2.2 \times 10^{44} h^{-2}$ erg s$^{-1}$ (0.3-3.5 keV band; Fabricant et al. 1986). We used the Las Campanas multi-fiber spectrograph (Shectman et al. 1992) to acquire spectra of 481 galaxies with $m_B \lesssim 19$ that lie within $1°$ of the cluster center (Abell 1958). We determine radial velocities with a cross-correlation method (Tonry & Davis 1979) or by fitting emission lines. A comparison with published data (Dressler & Shectman 1988) indicates that our velocities have $1\sigma$ observational uncertainties of $\sim 60$ km s$^{-1}$ and a zero point offset of $\lesssim 60$ km s$^{-1}$. Cluster membership is assigned to 341 galaxies in the redshift range of 0.045 to 0.063 (limits corresponding to the cluster mean velocity $16400 \pm (3 \times$ the cluster velocity dispersion of 900 km s$^{-1}$)). The data and detailed analyses will be presented in later work (Zabludoff 1995); here we focus on comparing the positions of cluster members on the sky with the cluster X-ray surface brightness distribution.

We acquired a 2322 sec exposure of A754 from the archives of the Position Sensitive Proportional Counter (PSPC) on the ROSAT satellite (Henry & Briel 1993). The X-ray emission from A754 lies within the central 40 arcmin of the $2°$ PSPC field, where the point response function is approximately Gaussian, and the flatfield and background levels are approximately constant. Because A754 is detected in both the hard and soft PSPC energy bands, we use X-ray photons in the broad 0.1-2.4 keV band in our analysis to include all the signal.

Plate 1 shows the smoothed X-ray map superimposed on the smoothed projected distribution of cluster galaxies, where each galaxy is equally weighted. We provide the details, including the smoothing scales, in the Plate 1 caption. The plate is an unbiased map of the optical light in the cluster; weighting the galaxies by their luminosity does not significantly change the features. The peak of the X-ray emission is at $(\alpha, \delta)_{2000} \sim (09^h\ 09^m\ 22^s, -09°\ 40'\ 11'')$. We adopt an uncertainty in these coordinates of $\sim 6$ arcsec, the nominal attitude accuracy for telescope pointing given by the ROSAT mission description. By comparing the coordinates of the field centers and of the x-ray/radio galaxy (Harris et al. 1984) in the southwestern corner of the plate, we estimate that the errors in the alignment of the X-ray and optical contours are $\lesssim 30$ arcsec.

Within the central $1h^{-1}$ Mpc, there is bimodal structure in the galaxy distribution. We assign a galaxy to either the southeast (SE) or northwest (NW) clump if the galaxy (1) lies within the clump's red optical contour in Plate 1 and (2) has a velocity within the limits corresponding to the clump mean velocity $\pm$ ($3 \times$ the clump velocity dispersion). Once memberships are assigned, we calculate the projected centroid of each clump. We estimate the uncertainty in the centroid position by throwing out each member, one at a time, and recalculating the centroid. The centroids vary during this procedure by $\lesssim 10$ arcsec. The SE galaxy clump lies at $(\alpha, \delta)_{2000} \sim (09^h\ 09^m\ 26^s, -09°\ 43'\ 49'')$ and contains 41 galaxies. The NW galaxy clump lies at $(\alpha, \delta)_{2000} \sim (09^h\ 08^m\ 27^s, -09°\ 37'\ 45'')$ and contains 27 galaxies. The two clumps are separated by 15.6 arcmin ($0.73h^{-1}$ Mpc) on the sky. The mean velocities and velocity dispersions of the galaxy clumps are $\overline{v}_{NW} = 16432 \pm 191$ km s$^{-1}$, $\sigma_{NW} = 970^{+156}_{-134}$ km s$^{-1}$ and $\overline{v}_{SE} = 16320 \pm 132$ km s$^{-1}$, $\sigma_{SE} = 830^{+130}_{-112}$ km s$^{-1}$ (where $\overline{v}$ and $\sigma$ are the bi-weight estimators of the line-of-sight location and scale; Beers et al. 1988).

## 3. DISCUSSION

A comparison of the X-ray and optical data in A754 yields three surprises. First, neither of the two major condensations of galaxies corresponds to the most prominent feature of the highly asymmetric X-ray surface brightness map in Plate 1. The X-ray peak is offset from the SE galaxy clump by a projected separation of 3.8 arcmin ($0.18h^{-1}$ Mpc), which is much greater than the relative uncertainty in their positions ($\lesssim 30$ arcsec). Second, emission from the X-ray peak is elongated perpendicular to the axis joining the SE and NW optical clumps. Third, a comparison of the galaxy clumps and the distribution of X-ray gas temperatures reveals that the two principal temperature components (Henriksen 1993; Henry & Briel 1995) are roughly aligned east-west along the axis connecting the optical clumps. Although the X-ray temperature of the eastern component ($\sim 5$ keV), which surrounds the SE galaxy clump, is consistent with that derived from $\sigma_{SE}$ (5.2 keV; $T_X = 10^{-3.22}\sigma^{1.35}$, Edge & Stewart 1991), the temperature of the western component is $\sim 13$ keV, much hotter than the 6.5 keV inferred from $\sigma_{NW}$.

The disturbed X-ray morphology is not due to the chance superposition of an unidentified X-ray source. The X-ray luminosity of the peak is $\sim 20$-$25\%$ that of the cluster. A single X-ray point source is ruled out by the elongation of the peak. The peak shape is not reproduced by a model of two point sources at any separation. Furthermore, no point sources are revealed in a deeper PSPC exposure of this field (Henry & Briel 1995). As for contamination from extended sources, there is no galaxy with $m_B < 19$ within the brightest contour of the X-ray peak. The two closest galaxies, which are cluster members, lie near the southern and northern ends of the elongated 2nd- and 3rd-brightest contours, respectively. The southern galaxy, which has no emission lines and is classified S0/D (Dressler 1980), is visually one of the brightest in the cluster ($m_B = 16.3$; Fabricant et al. 1986) and has a weak associated radio source (Harris et al. 1980). The northern galaxy ($m_B = 17.5$) is typed Irr (Dressler 1980); its spectrum has [OII] $\simeq 40$Å $>$ [OIII] $\simeq$ 10Å and so is consistent with a starburst. It is very unlikely that both galaxies have the X-ray luminosities ($> 10^{43}h^{-2}$ erg s$^{-1}$) and X-ray radii ($\sim 100h^{-1}$ kpc) necessary to account for the brightness and extent of the peak (Stocke et al. 1991; Fabbiano 1989). Visual inspection of the POSS plates finds no additional objects near the X-ray peak, except for an 8.9 magnitude F star (HD 78665) that makes a negligible contribution to the X-ray flux (Fabricant et al. 1986). Thus, we conclude that the X-ray emission of the peak originates from the intracluster gas.

What do the data for A754 imply about cluster evolution? We claim that there has been a recent collision of two subclusters in A754 and now discuss three observations that have led to this conclusion. First, the displacement between the galaxies, which we assume trace the dark matter, and the X-ray peak is consistent with that seen between the dark matter and the gas distributions during subcluster collisions in hierarchical clustering simulations (Evrard 1990b). During the collision, the models predict that the collisional X-ray gas is stripped from the free-streaming dark matter and galaxies. The gas is not in hydrostatic equilibrium with the dark matter potential for $\sim 1h^{-1}$ Gyr afterwards. The galaxies free-stream through the collision more easily than the gas, flying further than the gas from the point of impact after the clumps cross. It is suggestive that the SE galaxy clump and the X-ray peak lie in this configuration and that their separation is consistent with the predicted displacements ($\lesssim 0.5h^{-1}$ Mpc). The NW galaxy clump does not have an obvious X-ray counterpart, but one is implied by the east-west ellipticity of the cluster's central X-ray isophotes and by the hot X-ray temperature component in the west.

Second, the X-ray peak is elongated perpendicular to the collision direction. In simulations, this feature results from the compression of the gas after the intracluster media collide and persists for $\sim 1h^{-1}$ Gyr (Evrard 1990a,b; Roettiger et al. 1993). The high X-ray luminosity and relatively low X-ray temperature of the peak (compared to the hot western component) argue that it is a region of high gas density.

Third, there are at least two temperature components in A754 (Henry & Briel 1995), and the temperature of the western component is anomalous. The $\sim 12$ keV gas in the west, which is much hotter than inferred from the velocity dispersion of the galaxies in the NW clump, has been heated or even shocked beyond its equilibrium value, most likely by the compressive forces of the collision. Simulations predict that this type of heating occurs after the clumps collide (Roettiger et al. 1993), thereby supporting the picture in which the NW and SE galaxy clumps are not infalling for the first time, but have already crossed.

The excellent agreement of the data with the qualitative appearance and scale of the models suggests that the galaxy clumps in A754 have crossed in the plane of the sky during the last $\sim 1h^{-1}$ Gyr. We can test this picture with a simple model of two subclusters on linear orbits (Beers, Geller, & Huchra 1982). For the input age of the universe ($8.28h^{-1}$ Gyr; $q_0 = 0.1$), radial velocity difference between the galaxy clumps (112 km s$^{-1}$), clump projected separation ($R_p = 0.73h^{-1}$ Mpc), and total system mass ($M_{tot} = 4.8 \times 10^{14}\ M_\odot$), there is a bound, outgoing solution of the motion equations in which the angle between the collision axis and the sky is 3.1 degrees, the clump separation is $0.73h^{-1}$ Mpc ($\simeq R_p$), the relative clump velocity is 2071 km s$^{-1}$, and the time elapsed since the clumps crossed is $0.2h^{-1}$ Gyr. The system mass $M_{tot}$, which is the virial mass estimated from the 57 cluster members within the circle whose origin is the centroid of the two clumps and whose diameter is $R_p$, is consistent with the sum of the virial masses of the two clumps. Changing $M_{tot}$ by a factor of two does not affect the character of the solution. Because this model assumes that (1) the subclusters are point masses, (2) the orbits are linear, (3) the influence of the perturbed intracluster medium is negligible, and (4) the subcluster masses are constant, we use the model only to demonstrate the plausibility of our particular collision scenario.

Plate 1 also provides insight into why cD galaxies in some clusters (Beers & Geller 1983; Zabludoff et al. 1990) do not lie on the projected, optical cluster center or at the bottom of the cluster potential well. The cD galaxy in A754 (Dressler 1980) is within 1.2 arcmin of the centroid of the NW galaxy clump and has the same velocity as the clump mean velocity. The galaxy is therefore at rest in a local density maximum, but is displaced relative to the projected, optical center of A754 as a whole. For subclusters on linear orbits in the plane of the sky, the cD's offset from the cluster center is maximized and its peculiar motion with respect to the cluster mean velocity minimized. We can imagine a less fortuitous configuration in which the collision axis is rotated 90 degrees into the line-of-sight. In this case, both the bimodality of the galaxy distribution and the displacement of the cD could only be detected in the distribution of galaxy radial velocities. The collision signatures would be nearly impossible to observe. Thus, any determination of the fraction of clusters with substructure or with recent subcluster collisions is a lower limit.

A more complete understanding of the evolution of the subcluster collision requires n-body/hydrodynamical simulations that reproduce the specific optical and X-ray properties of A754. New ASCA X-ray temperature maps with high spatial resolution may reveal the existence of a shock and the details of its structure. For now, we conclude that estimating cluster X-ray masses by assuming isothermality and hydrostatic equilibrium is perilous. In A754, the derived mass, assumed directly proportional to temperature, is overestimated by $\sim 50\%$ because the gas is heated beyond its equilibrium temperature (Henry & Briel 1993). This effect may contribute to the discrepancies (Babul & Miralda-Escudé 1994) sometimes found between cluster X-ray masses and the masses derived from gravitational lens models.

How common are collisions and what are the implications for cosmology? In hierarchical models, the rate at which clusters accrete mass is especially sensitive to the mean mass density parameter $\Omega_0$ (cf. Richstone et al. 1992; Evrard et al. 1993; Lacey & Cole 1993). Therefore, we can place stringent limits on $\Omega_0$ by determining the number of nearby clusters in which subclusters have recently collided. The observed fraction of clusters with substructure provides some estimate of the accretion frequency, but depends critically on (1) whether the apparent substructure is real or projected and (2) the time elapsed before the subclusters collide and virialize. This relaxation timescale in turn depends on the energy and eccentricity of the subcluster orbits and on the distribution of mass within the cluster. Because the relaxation timescale is poorly constrained, it is not possible to know whether the substructure observed today reflects predominantly recent or ancient accretion. This distinction is critical for the measurement of $\Omega_0$ because the accretion rate is a strong function of lookback time. For example, the current fraction of clusters with substructure for an $\Omega_0 = 0.2$ model changes from 5% to 40% depending on whether the accretion occurred within the last $\sim 2h^{-1}$ or $\sim 4h^{-1}$ Gyr (Lacey & Cole 1993). As we show in A754, we can avoid most of these uncertainties by using the physics of both the gas and galaxies to directly identify clusters in which subclusters have collided in the last $\sim 1h^{-1}$ Gyr.

We stress that A754 is unusual only because it has been examined on scales where the effects of subcluster collisions are most visible. Similar displacements between the galaxies and peak of the X-ray emission are observed in a subcluster in Coma (Mellier et al. 1988; Briel et al. 1992) and in Virgo (Binggeli et al. 1993; Bohringer et al. 1994). It is noteworthy that Coma and Virgo are the only two other clusters for which there are as many measured galaxy velocities as for A754. On-going collisions might also be the cause of inconsistencies between the global kinematics of the gas and of the galaxies in Perseus (Gorenstein et al. 1978; Cowie, Henriksen, & Mushotzky 1987) and in Abell 2256 (Fabricant, Kent, & Kurtz 1989). A detailed comparison of the optical and X-ray properties of a large, well-defined sample of clusters will reveal the incidence of such collisions and will place an improved constraint on cosmological models.

We thank Marc Postman for providing digitized POSS images, Huan Lin and Steve Shectman for advice concerning the reduction of multi-fiber spectrograph data, and Anand Sivaramakrishnan for his photography expertise. Dan Fabricant, Pat Henry, Gus Evrard, Ian Smail, Megan Donahue, John Mulchaey, Simon White, and Richard Mushotzky contributed helpful information and comments. AIZ acknowledges support from a Carnegie Fellowship and from the Dudley Observatory. DZ acknowledges partial support provided by NASA through grant HF-1027.01-91A from STScI, which is operated by AURA, Inc. under NASA contract NAS5-26555.

## Plate Caption

**Plate 1:** The smoothed distribution of X-ray emitting gas (white contours) superimposed on the smoothed distribution of cluster galaxies in A754. The scale of the plot is $46.8h^{-1}$ kpc arcmin$^{-1}$ ($q_0 = 0.1$). Note that neither of the two major clumps of galaxies (bordered by red) lies on the extended peak of the X-ray emission. The distribution of 201 galaxies is smoothed with a Gaussian of 1.23 arcmin ($\sim 60h^{-1}$ kpc) FWHM, a scale smaller than that of giant galaxy halos (Zaritsky & White 1994). The faintest greyscale level corresponds to individual galaxies. The PSPC image is smoothed with a Gaussian of 30 arcsec FWHM; the resulting map has a resolution of $\sim 40$ arcsec ($\sim 30h^{-1}$ kpc) determined from the FWHM of the X-ray point source $\sim 8$ arcmin north of the X-ray peak. The background level is $7.0 \pm 0.1 \times 10^{-4}$ counts s$^{-1}$ arcmin$^{-2}$. The lowest contour level is four times the background. The highest level plotted is 0.04 counts s$^{-1}$ arcmin$^{-2}$ ($57 \times$ the background). The difference between each contour is $2.9 \times 10^{-3}$ counts s$^{-1}$ arcmin$^{-2}$. The bright X-ray source in the southwest is associated with a radio galaxy in a group of galaxies at the same redshift as A754 (Harris, Constain & Dewdney 1984).

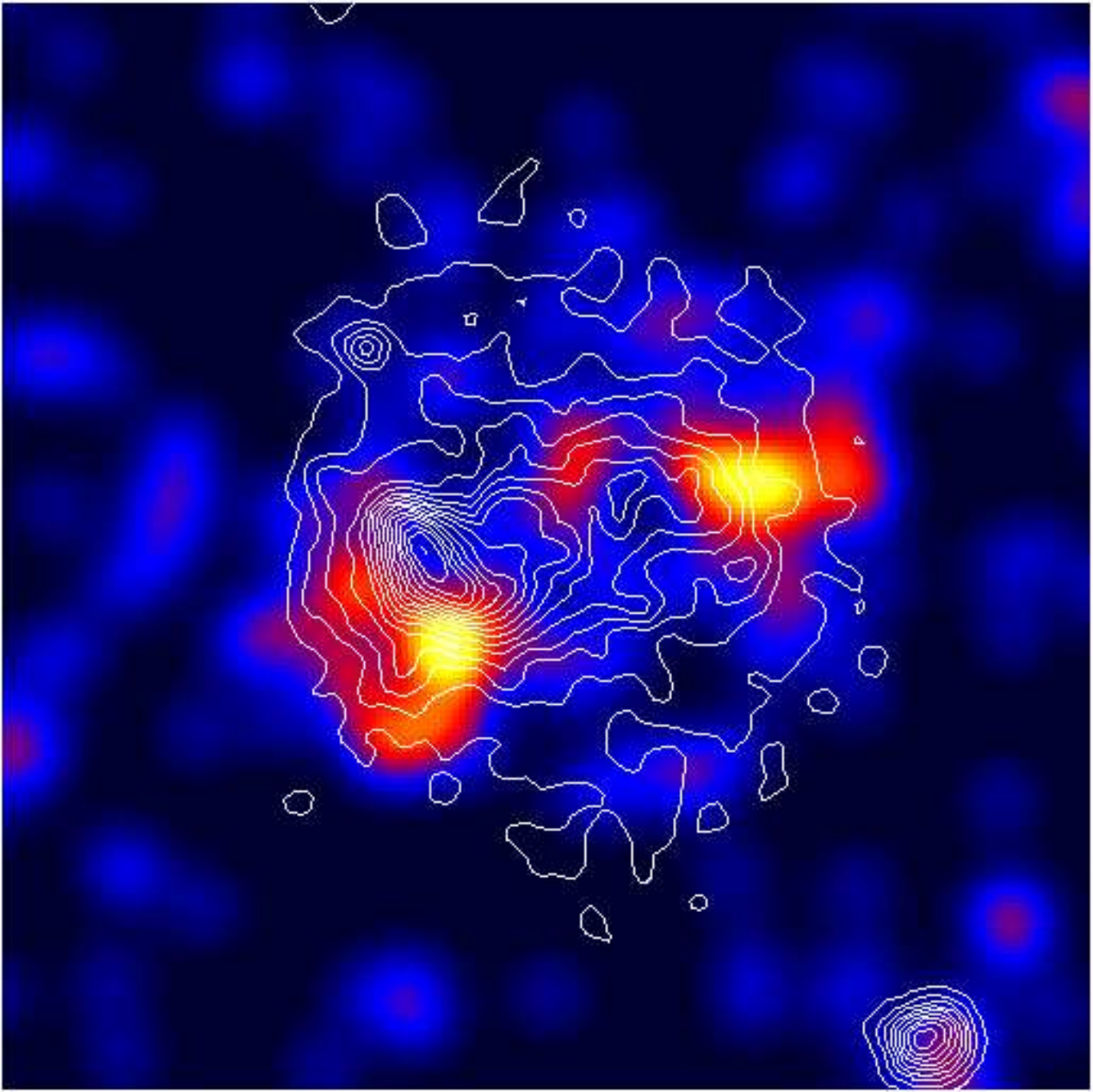